\documentclass[conference,a4paper]{IEEEtran}
\usepackage{amssymb}
\usepackage{amscd}
\usepackage{amsmath}
\usepackage{epsfig}
\usepackage{amsthm}
\usepackage{graphics}
\usepackage{psfrag}
\usepackage{rotating}
\usepackage{amsmath}
\usepackage{amsfonts}
\usepackage{url}
\usepackage{color}
\usepackage{epstopdf}
\usepackage{amsthm}
\usepackage{tikz}
\usepackage{mathtools}
\usepackage{multirow}
\usetikzlibrary{arrows,shapes,decorations,backgrounds}
\usepackage[final]{pdfpages}
\usepackage{enumitem}
\usepackage{multicol}
\usepackage[nolist]{acronym}
\usepackage{subcaption}

\newtheorem{remark}{Remark}

\newfont{\bbb}{msbm10 scaled 700}

\newfont{\bb}{msbm10 scaled 1100}
\newcommand{\CC}{\mbox{\bb C}}

\newcommand{\RR}{\mbox{\bb R}}

\newcommand{\EE}{\mbox{\bb E}}

\newcommand{\av}{{\bf a}}
\newcommand{\bv}{{\bf b}}

\newcommand{\fv}{{\bf f}}
\newcommand{\gv}{{\bf g}}
\newcommand{\hv}{{\bf h}}

\newcommand{\rv}{{\bf r}}

\newcommand{\wv}{{\bf w}}

\newcommand{\xv}{{\bf x}}
\newcommand{\yv}{{\bf y}}

\newcommand{\Am}{{\bf A}}
\newcommand{\Bm}{{\bf B}}

\newcommand{\Gm}{{\bf G}}
\newcommand{\Hm}{{\bf H}}
\newcommand{\Id}{{\bf I}}

\newcommand{\Xm}{{\bf X}}

\newcommand{\Fc}{{\cal F}}

\newcommand{\Pc}{{\cal P}}

\newcommand{\Tc}{{\cal T}}

\newcommand{\thetav}{\hbox{\boldmath$\theta$}}

\newcommand{\Psim}{\hbox{\boldmath$\Psi$}}

\renewcommand{\arg}{{\hbox{arg}}}

\renewcommand{\Re}{{\rm Re}}

\newcommand{\eqdef}{\stackrel{\Delta}{=}}

\newcommand{\Pav}{P_{\rm avg}}
\newcommand{\Na}{N_a}
\newcommand{\Nrf}{N_{\rm rf}}
\renewcommand{\H}{{\scriptscriptstyle\mathsf{H}}}

\newcommand\blfootnote[1]{%
	\begingroup
	\renewcommand\thefootnote{}\footnote{#1}%
	\addtocounter{footnote}{-1}%
	\endgroup
}

\begin{document}

\begin{acronym}
	\acro{AWGN}{additive white Gaussian noise}
	\acro{MIMO}{multiple-input multiple-output}
	\acro{OTFS}{orthogonal time frequency space}
	\acro{SNR}{signal-to-noise ratio}
	\acro{mmWave}{millimeter wave}
	\acro{PL}{pathloss}
	\acro{ML}{maximum likelihood}
	\acro{V2X}{vehicle-to-everything}
	\acro{OFDM}{orthogonal frequency division multiplexing}
	\acro{FMCW}{frequency modulated continuous wave}
	\acro{LoS}{line-of-sight}
	\acro{ISFFT}{inverse symplectic finite Fourier transform}
	\acro{SFFT}{symplectic finite Fourier transform}
	\acro{HPBW}{half-power beamwidth}
	\acro{ULA}{uniform linear array}
	\acro{CRLB}{Cram\'er-Rao Lower Bound}
	\acro{RF}{radio frequency}
	\acro{BF}{beamforming}
	\acro{RMSE}{root MSE}
	\acro{AOA}{angle of arrival}
	\acro{ISI}{inter-symbol interference}
\end{acronym}

\title{Joint Radar Target Detection and Parameter Estimation with MIMO OTFS}

\author{
	\IEEEauthorblockN{Lorenzo Gaudio$^{1}$, Mari Kobayashi$^{2}$, Giuseppe Caire$^{3}$, Giulio Colavolpe$^{1}$}
	\IEEEauthorblockA{$^{1}$University of Parma, Italy \\
		$^{2}$Technical University of Munich, Munich, Germany \\
		$^{3}$Technical University of Berlin, Germany\\ 
		Emails: lorenzo.gaudio@studenti.unipr.it, mari.kobayashi@tum.de, giulio.colavolpe@unipr.it, caire@tu-berlin.de }
	}

\maketitle

% \thanks{The work of Lorenzo Gaudio, Giuseppe Caire, and Giulio Colavolpe is supported by Fondazione Cariparma, under the TeachInParma Project. The work of Mari Kobayashi is supported	by an Alexander von Humboldt Research Fellowship.}

\begin{abstract}
Motivated by future automotive applications, we study the joint target detection and parameter estimation problem using orthogonal time frequency space (OTFS), a digital modulation format robust to time-frequency selective channels. Assuming the transmitter is equipped with a mono-static MIMO radar, we propose an efficient maximum likelihood based approach to detect targets and estimate the corresponding delay, Doppler, and angle-of-arrival parameters. 
In order to reduce the computational complexity associated to the high-dimensional search, our scheme proceeds in two steps, i.e., target detection and coarse parameter estimation followed by refined parameter estimation. Interestingly, our numerical results demonstrate that the proposed scheme is able to identify multiple targets if they are separated in at least one domain out of three (delay, Doppler, and angle), while achieving the Cram\'er-Rao lower bound for the parameter estimation. 

%Nowadays many automotive systems address the problem of joint radar and communication. In order to provide a system able to jointly perform common radar tasks together with the transmission of useful information to a target receiver, we focus our attention on OTFS, being a digital modulation format not sensitive to Doppler and delay shifts, and consider the problem of target detection and joint delay, Doppler, and angle of arrival estimation when the transmitter and the radar receiver are co-located and equipped with an antenna array. We show that multiple targets are well identified by the proposed joint radar and communication system if separable in at least one domain out of three (delay, Doppler, or angle). Moreover, the application of a super-resolution maximum likelihood approach for both target detection and radar parameters estimation pushes the errors towards the theoretical benchmark, i.e., the Cram\'er-Rao lower bound. In order to cope with the computational complexity of the maximum likelihood estimator, we propose an efficient multiple-step refinement scheme, such that a first coarse estimation and detection of the targets is followed by a finer evaluation of the radar parameters. The analytical derivation the joint detection/estimation algorithm are supported by numerical simulations, highlighting the fact that OTFS, the communication-oriented waveform analyzed, is a candidate solution for the aforementioned problem.

\end{abstract}
	
\section{Introduction}
\blfootnote{The work of Lorenzo Gaudio, Giuseppe Caire, and Giulio Colavolpe is supported by Fondazione Cariparma, under the TeachInParma Project. This research benefits from the HPC (High Performance Computing) facility of the University of Parma, Italy.}The \ac{MIMO} radar has been extensively studied and has been shown to improve the resolution, i.e., the ability to distinguish multiple targets, thanks to an additional spatial dimension (see, e.g., \cite{li2008mimo}). A careful design of \ac{BF}, or power allocation along angular directions, is crucial to achieve accurate radar detection and parameter estimation performance. This is particularly relevant to automotive radar \cite{patole2017automotive} operating over \ac{mmWave} frequency bands as high propagation loss must be compensated by proper \ac{BF}, or more generally beam alignment both at transmitter and receiver sides (see e.g. \cite{song2018scalable} and references therein). 
Note that in a mono-static radar system with co-located transmitter and receiver, transmit and receive antennas are calibrated such that their beam patterns are consistent, i.e., ``look in the same direction''.  Moreover, \ac{BF} at the radar transmitter might be adaptive, depending on different operating phases (see e.g., \cite{friedlander2012transmit,niesen2019joint} and references therein). Namely, the transmitted power shall be allocated to wider angular sectors during a target detection/search phase, while narrow and distinct beams, each focused on the detected target, shall be used to minimize  ``multi-target'' interference
in a tracking phase \cite{li2008mimo,richards2014fundamentals,bar1995multitarget}. During the target detection phase, a non-trivial tradeoff appears. On one hand, a wider angular sector coverage enables to detect multiple targets if the received backscattered power is high enough. On the other hand, a more directional \ac{BF} grants a higher received \ac{SNR}, at the cost of a time-consuming search over narrower angular sectors. 

%Automotive radar operates typically over \ac{mmWave} frequency bands and high propagation loss must be compensated by proper beamforming, or more generally beam alignment both at transmitter and receiver sides (see e.g. \cite{song2018scalable} and references therein). 
%Additionally, having the possibility to choose among practically any carrier frequency, the system can be setup to match the desired level of accuracy during %radar detection and estimation processes, i.e., the radar resolution, also thanks to an almost-free choice of the bandwidth, whose availability generally %increases with the carrier frequency.
%associated to high- and very-high carrier frequencies, e.g., \ac{mmWave} \cite{swindlehurst2014millimeter,pi2011introduction}, 

As an extension of our previous work \cite{gaudio2019effectiveness}, this paper studies the joint problem of target detection and parameter estimation with a \ac{MIMO} mono-static radar adopting \ac{OTFS}, i.e., a multi-carrier communication waveform recently proposed in \cite{hadani2017orthogonal}. 
The use of communication waveforms for radar has been motivated by the joint radar and communication paradigm, where two functions are implemented by sharing the same resources and the same waveform (see e.g. \cite{dokhanchi2019mmwave,zheng2019radar,hassanien2019dual} and references therein). In particular, delay and Doppler estimation based on \ac{OTFS} has been also considered in \cite{raviteja2019orthogonal}, where the authors propose a matched filter based approach to estimate the parameters within discrete delay-Doppler grids. In this work, we assume that the coverage is wide enough such that multiple targets could be located at different angles within the same beam, contrary to the time-consuming beam sweeping considered in \cite{gaudio2019effectiveness}. Under this setup, we aim to find an efficient method for target detection and delay, Doppler, and \ac{AOA} estimation under practical \ac{OTFS} system constraints related to the underlying physical channel, explained within the paper. 

Our main contributions are two-fold: i) we propose an efficient two-step approach for joint target detection and four-dimensional parameter estimation. 
In order to reduce the computational complexity associated to the high-dimensional search, our \ac{ML} based scheme first performs coarse estimation and target detection, and then refines further the parameter estimation; ii) simulation results demonstrate that our proposed approach can identify multiple targets if they are separated in at least one domain out of three (delay, Doppler, and angle), while achieving the \ac{CRLB} for the parameter estimation. 

The paper is organized as follows. In Section \ref{sec:phy-model}, we introduce the physical model (of the channel) and the \ac{OTFS} input-output relation. Section \ref{sec:Joint-Detection-Param-Estimation} describes our proposed joint detection and parameters estimation algorithm, together with the definition of the \ac{CRLB}. After showing the simulation results in Section \ref{sec:Sim-Results}, Section \ref{sec:Conclusion} concludes the paper. 

\section{Physical model}\label{sec:phy-model}

%We consider joint radar detection and parameter estimation in a system that uses the same waveform and transmitter for communication and radar purposes. We assume that a transmitter sends a digitally modulated waveform to communicate information to a given receiver, while using the backscattered wave for radar detection and estimation. The radar receiver is co-located with the transmitter, i.e., the radar detection and estimation is performed at the same transmitter location (mono static, full duplex). The communication aspects of the problem enter the motel only in the selection of the waveform, which is OTFS in our case. For more details on the data transmission performance of OTFS please refer to [put refs on OTFS, including our own papers]. 

We consider joint radar detection and parameter estimation in a system operating over a channel bandwidth $B$ at the carrier frequency $f_c$.  We assume 
that a transmitter is equipped with a mono-static \ac{MIMO} radar with $\Na$ antennas and operating in full duplex.\footnote{Full-duplex operations can be achieved with sufficient isolation between the transmitter and the (radar) detector and possibly interference analog pre-cancellation in order to prevent the (radar) detector saturation \cite{sabharwal2014band}.} Wide angular sectors are illuminated by the transmit beam and the receiver processes the backscattered signal to identify the presence of targets within the beam, together with the estimation of parameters of interest such as range, velocity, and angular position. 
% A possibly successive phase, i.e., targets tracking \cite{niesen2019joint}, in which the transmitter \ac{BF} illuminates with one beam per-target the detected targets of the first phase, to convey a message while estimating the parameters of interest listed before. The latter case has been investigated in \cite{gaudio2019effectiveness}. 

In order to take advantage of the available bandwidth, automotive radar systems typically operate at \ac{mmWave} carrier frequencies \cite{patole2017automotive}. At \ac{mmWave}, although the number of antennas may be very large, the number of \ac{RF} chains is limited due to the difficulty of implementing a full per-antenna \ac{RF} chain (including A/D conversion, modulation, and amplification) in a small form factor and highly integrated technology, for large signal bandwidths. Full digital \ac{BF} is thus precluded, while hybrid digital-analog \ac{BF} schemes is typically adopted (see, e.g., \cite{song2019fully} and references therein). Targeting possible \ac{mmWave} automotive applications, we consider a number of \ac{RF} chains smaller than the number of antennas, i.e., $\Nrf\leq \Na$. %, while the extension to a full digital \ac{BF} architecture is straightforward.

We consider a point target model such that each target can be represented through its \ac{LoS} path only \cite{kumari2018ieee,nguyen2017delay}. By letting $\phi\in [-\frac{\pi}{2}, \frac{\pi}{2}]$ be the steering angle, the transmitter and receiver arrays are given by $\av(\phi)$ and $\bv(\phi)$, where $\bv(\phi)=(b_1(\phi), \dots, b_{N_a}(\phi))\in \CC^{N_a}$ denotes the uniform linear array response vector of the radar receiver, given by
\begin{align}
	b_n(\phi) &= e^{j(n-1)\pi \sin(\phi)},\;\; n=1,\dots, N_a\,,
\end{align}  
and $\av(\phi)$ is defined similarly (at the radar transmitter). The channel is modeled as a $P$-tap time-frequency selective channel of dimension $\Na\times\Na$ given by \cite{vitetta2013wireless}
\begin{align}\label{eq:Channel}
	\Hm (t, \tau) =  \sum_{p=1}^P h_p \bv(\phi_p) \av^\H (\phi_p)\delta(\tau-\tau_p)  e^{j2\pi \nu_p t}\,,
\end{align}
where $(\cdot)^\H$ denotes the Hermitian (conjugate and transpose) operation, $P$ is the number of targets, $h_p$ is a complex channel gain including the \ac{PL} of the path component, $\nu_p = \frac{2v_p f_c}{c}$, $\tau_p=\frac{2r_p}{c}$ denote the corresponding round-trip Doppler shift, delay associated to the $p$-th target, respectively. %, $2v_p$, and $2r_p$ 

%By denoting with $\fv_i^{\Na\times1}$ a generic complex \ac{BF} vector and with $s_i(t)$ the signal sent by the $i$-th \ac{RF} chain, the continuous transmit signal is given by
%\begin{align}
%	\xv (t) = \sum_{i=1}^{\Nrf} \fv_i s_i(t)=\Fm\sv(t)\,,
%\end{align}
%where $\trace(\EE[\xv(t) \xv(t)^\H]) \leq \Pav$, $\fv_i$ is the $i$-th column of the \ac{BF} matrix $\Fm$, of dimension $\Na\times\Nrf$ and detailed afterwords, and $\sv(t)$ collects the entries $s_i(t)$, for $i=1,\dots,\Nrf$. The signal $\sv(t)$ is thus sent to all antennas.

\subsection{OTFS Input Output Relation}\label{subsec:OTFS-Input-Output}
We consider \ac{OTFS} with $M$ subcarriers of bandwidth $\Delta f$ each, such that the total bandwidth is given by $B=M\Delta f$. We let $T$ denote the symbol time, and the \ac{OTFS} frame duration is $NT$, imposing $T\Delta f=1$. By following the standard derivation of the input-output relation of \ac{OTFS} (see, e.g., \cite{hadani2017orthogonal,gaudio2019effectiveness}), data symbols $\left\{x_{k,l}\right\}$, for $k=0,\dots,N-1$ and $l=0,\dots,M-1$ belonging to any constellation, are arranged in an $N\times M$ two-dimensional grid referred to as the Doppler-delay domain, i.e., $\Gamma=\left\{\left({k}/{NT},{l}/{M\Delta f}\right)\right\}$ for $k=0,\dots,N-1$ and $l=0,\dots,M-1$. 
% In order to send the block of symbols $\{x_{k,l}\}$, 
The transmitter first applies the \ac{ISFFT} to convert data symbols $\{x_{k,l}\}$ into a block of samples $\{X[n, m]\}$ in the dual domain, referred to as the time-frequency domain, thus
\begin{equation}\label{eq:x-to-X}
	X[n,m]=\sum_{k=0}^{N-1}\sum_{l=0}^{M-1}x_{k,l}e^{j2\pi\left(\frac{nk}{N}-\frac{ml}{M}\right)}, 
\end{equation}
for $n = 0, \ldots, N-1$ and $m = 0, \ldots, M-1$. 
Then, it generates the continuous-time signal
\begin{align}\label{eq:Tx-Signal-General}
	s(t)  = \sum_{n=0}^{N-1} \sum_{m=0}^{M-1} X[n, m] g_{\rm tx}(t-nT) e^{j 2\pi m \Delta f(t-nT)}\,,
\end{align}
where $X[n, m]$ denotes the symbol sent at time $n$ over subcarrier $m$, satisfying the average power constraint $\EE[|X[n ,m]|^2]\leq {\Pav}/{N_a}$. 
For simplicity, we consider that the same symbol stream $\{X[n,m]\}$ is repeated over all 
\ac{RF} chains. After transmission over the channel defined in \eqref{eq:Channel}, the continuous received signal without noise including \ac{BF} is
\begin{align}\label{eq:Received-Signal-First}
	\rv(t) = \sum_{p=0}^{P-1} h_p \bv(\phi_p) \av^\H(\phi_p)\fv_{\mathrm{BF}} s(t-\tau_p) e^{j2\pi \nu_pt}\,,
\end{align}
% where $\Fm$ is a generic beamforming matrix of dimension $\Na\times\Nrf$, with columns $\fv_i^{\Na\times1}$ corresponding to complex \ac{BF} vectors. 
where $\fv_{\mathrm{BF}}$ is a generic \ac{BF} vector of dimension $\Na\times1$. Typical \ac{BF} designs can be found, e.g., in \cite{friedlander2012transmit,fortunati2020mimoRadar}.
% In \eqref{eq:Received-Signal-First}, we also exploit the single stream condition, such that the column vector $\mathbf{1}_{\Nrf\times 1}$ of ones appears. Thus, without loss of generality we can focus on the case of a single \ac{RF} chain collecting all the transmitted power and a \ac{BF} vector given by $\fv_{\mathrm{BF}} = \Fm\cdot\mathbf{1}_{\Nrf\times1}$. By focusing on target detection, the assumption of a single \ac{RF} chain, together with the definition of a \ac{BF} vector exploring a wide angular sector, holds.

The output of the receiver filter-bank adopting a generic receive shaping pulse $g_{\rm rx}(t)$ is given in \eqref{eq:y-received},
\begin{figure*}
	\begin{align}\label{eq:y-received}
	\yv(t&, f) = \int  \rv(t')g^*_{\rm rx} (t'-t) e^{-j2\pi ft'} dt'= \int_{t'} g^*_{\rm rx} (t'-t) \sum_{p=0}^{P-1} h_p \bv(\phi_p) \av^\H(\phi_p)\fv_{\mathrm{BF}}s(t'-\tau_p) e^{j2\pi \nu_pt'} e^{-j2\pi ft'} dt'\nonumber\\
	&= 
	%\sum_{n'=0}^{N-1}\sum_{m'=0}^{M-1} \sum_{p=0}^{P-1} 
	\sum_{p,n',m'}h_p \bv(\phi_p)\av^\H(\phi_p) \fv_{\mathrm{BF}} X[n', m']\int_{t'} g^*_{\rm rx} (t'-t)  g_{\rm tx}(t'-\tau_p-n'T) e^{j 2\pi m' \Delta f(t'-\tau_p-n'T)} e^{j2\pi (\nu_p-f) t'} dt'
	\end{align} 
\end{figure*}
and, by sampling at $t=nT$ and $f=m\Delta f$, we obtain
\begin{align}
	\yv[n, m] &=\yv(t, f)|_{t=n T,f=m\Delta f} \nonumber\\
	 &=\sum_{n'=0}^{N-1} \sum_{m'=0}^{M-1} X[n', m'] \hv_{n,m}[n', m']\,,
\end{align}
where the time-frequency domain channel $\hv_{n,m}[n', m']$ is given in \eqref{eq:time-frequency-h},
\begin{figure*}
	\vspace{-0.5cm}
	\begin{align}\label{eq:time-frequency-h}
	\hv_{n,m}[n', m']= \sum_{p=0}^{P-1} h'_p \bv(\phi_p) \av^\H(\phi_p)\fv_{\mathrm{BF}}C_{g_{\rm tx},g_{\rm rx}}((n-n')T -\tau_p, (m-m') \Delta f-\nu_p) e^{j 2\pi n' T \nu_p} e^{-j2\pi m \Delta f  \tau_p}
	\end{align}
\end{figure*}
by defining the cross ambiguity function $C_{u, v}(\tau, \nu) \eqdef \int_{-\infty}^{\infty} u(s) v^*(s-\tau) e^{-j 2\pi \nu s} ds$ as in \cite{matz2013time}, letting $h'_p = h_p e^{j 2\pi \nu_p \tau_p}$, and imposing the term $e^{-j2\pi mn'\Delta f T}=1$, $\forall n',m$, which is always true under the hypothesis $T\Delta f=1$.
Since $X[n, m]$ is generated via \ac{ISFFT}, the received signal in the delay-Doppler domain is obtained by the application of the \ac{SFFT}
\begin{align}
	\yv[k,l]\!=\!\sum_{n,m}\frac{\yv[n,m]}{NM}e^{j2\pi\left(\frac{ml}{M}-\frac{nk}{N}\right)}\!\!=\!\!\sum_{k', l'} x_{k',l'} \gv_{k,k'}\left[l,l'\right],
\end{align}
where the \ac{ISI} coefficient of the Doppler-delay pair $\left[k',l'\right]$ seen by sample $\left[k,l\right]$ is given by
\begin{equation}
	\gv_{k,k'}\left[l,l'\right]=\sum_p h_p' \bv(\phi_p)\av^\H(\phi_p)\fv_{\mathrm{BF}}\Psi^p_{k, k'}[l,l']\,,
\end{equation}
with $\Psi^p_{k, k'}[ l, l']$ defined in \eqref{eq:Psi-Mat}.
\begin{figure*}
	\vspace{-0.5cm}
	\begin{align}\label{eq:Psi-Mat}
	\Psi^p_{k, k'}[ l, l'] =\!\!\! \sum_{n, n', m, m'}\!\!\! \frac{C_{g_{\rm rx}, g_{\rm tx}}((n-n')T -\tau_p, (m-m') \Delta f-\nu_p)}{NM}  e^{j 2\pi n' T \nu_p} e^{-j2\pi m \Delta f  \tau_p}e^{j 2\pi \left(\frac{n'k'}{N}- \frac{m'l'}{M}\right)}e^{-j 2\pi\left(\frac{nk}{N}-\frac{ml}{M}\right)}
	\end{align}
	\noindent\rule{\textwidth}{0.4pt}
\end{figure*}
Note that a simplified version of $\Psi^p_{k, k'}[ l, l']$ obtained by approximating the cross ambiguity function can be found in \cite{gaudio2019effectiveness}.

At this point, let's define 
\begin{align}
	\Gm_p(\tau_p, \nu_p,\phi_p)\triangleq\left(\bv\left(\phi_p\right)\av^\H(\phi_p)\fv_{\mathrm{BF}}\right)\otimes\Psim^p\,,
\end{align} 
where $\otimes$ is the Kronecker product,\footnote{Note that $\av^{X\times1}\otimes \Am^{Y\times Z}=\Bm^{XY\times Z}$.} as the $\Na NM\times NM$ matrix obtained by multiplying $\Psim^p$ by a different coefficient of $\left(\bv\left(\phi_p\right)\av^\H(\phi_p)\fv_{\mathrm{BF}}\right)$. Thus, by stacking $\Xm$ into a $NM$-dimensional vector $\xv$ and defining an output vector $\yv$ of dimension $NM\Na\times 1$, the received signal in the presence of noise is given by
\begin{align}
	\yv = \sum_{p=0}^{P-1} \left[h'_p\Gm_p(\tau_p, \nu_p,\phi_p)\right]\xv  + \wv\,,
\end{align}
where $\wv$ denotes the \ac{AWGN} vector with independent and identically distributed entries of zero mean and variance $\sigma_w^2$. The problem reduces to detect $P$ targets and estimate the $4P$ associated parameters (complex channel coefficient, Doppler, delay, and angle) from the $\Na MN$-dimensional received signal.

\section{Joint Detection and Parameters Estimation}\label{sec:Joint-Detection-Param-Estimation}
We wish to estimate the set of four parameters $\thetav = \{h'_p, \phi_p, \tau_p, \nu_p\}\in \Tc^P$, with $\Tc=  \CC \times \RR \times \RR \times \RR$. We define the \ac{ML} function as 
\begin{align}\label{eq:497}
	l(\yv| \thetav, \xv) & = \left|\yv - \sum_p h'_p \Gm_p \xv  \right|^2, %\nonumber \\
	%&= \yv^\H \yv -\yv^\H \sum_p h'_p \Gm \xv - \sum_p h_p'^* \xv^\H\Gm^\H \yv  + \xv^\H \left(\sum_p h'_p \Gm\right)^\H  \left(\sum_q h'_q \Gm_q \right) \xv\,,
\end{align}
where we use the short hand notation $\Gm_p\triangleq\Gm(\tau_p, \nu_p,\phi_p)$.\footnote{Operator $\left|\cdot\right|$ denotes the absolute value $\left|x\right|$ if $x\in\RR$, or the cardinality (number of elements) of a discrete set, i.e., $\left|\Fc\right|$, if $\Fc$ is a discrete set.} The \ac{ML} solution is given by 
\begin{align}
	\hat{\thetav} = \arg\min_{\thetav \in \Tc^P}  l(\yv| \thetav, \xv).
\end{align}
For a fixed set of $\{ \phi_p, \tau_p, \nu_p\}$, the \ac{ML} estimator of $\{h'_p\}$ is given by solving the following set of equations 
\begin{align}\label{eq:459}
	% \xv^\H\Gm^\H(\tau_p, \nu_p, \phi_p) \left(\sum_{q=0}^{P-1} h'_q  \Gm(\tau_q, \nu_q, \phi_q) \right) \xv &=  \xv^\H  \Gm^\H(\tau_p, \nu_p, \phi_p) \yv, \;\;\; p=0,\dots, P-1. 
	\xv^\H\Gm^\H_p\left(\sum_{q=0}^{P-1} h'_q  \Gm_q\right) \xv &=  \xv^\H  \Gm^\H_p \yv, \;\;\; p=0,\dots, P-1.  
\end{align}
By plugging \eqref{eq:459} into \eqref{eq:497}, it readily follows that minimizing $ l(\yv| \thetav, \xv)$ reduces to maximize
\begin{align}
	l_2(\yv| \thetav, \xv) &= \sum_p h'_p  \yv^\H\Gm_p \xv 
 	%&=  \sum_p  \frac{ \xv^\H  \Gm^\H \left[\yv   - \left(\sum_{q\neq p}h'_q  \Gm_q \right) \xv \right]  \yv^H\Gm \xv}{\xv^\H \Gm^\H \Gm \xv} \\
 	\nonumber\\
	&= \sum_p S_p(\tau_p, \nu_p, \phi_p) - I_p( \{h'_q\}_{q\neq p}, \thetav)\,,
\end{align}
where $S_p(\tau_p, \nu_p, \phi_p)$ and  $I_p( \{h'_q\}_{q\neq p},\thetav)$ ($S_p$ and $I_p$ in short hand notation) denote the useful signal and the interference term for target $p$, given respectively by
\begin{align}
	S_p  =  \frac{| \yv^\H\Gm_p \xv|^2}{| \Gm_p \xv|^2}\,,
\end{align}
\begin{align}
	 I_p =\frac{\left(\yv^\H\Gm_p \xv\right) \xv^\H  \left(\Gm^\H_p\sum_{q\neq p}h'_q  \Gm_q\right)  \xv}{| \Gm_p \xv|^2}\,.
\end{align}

The algorithm for joint target detection and parameter (\ac{AOA}, Doppler, and delay) estimation is summarized in the following.
\begin{enumerate}
	\item \textbf{\textit{Detection --- (AOA, Doppler, Delay) Coarse Estimation:}}\\ 
	We look for a set of possible targets 
	\begin{align}\label{eq:3D-Search}
		\Pc=\left\{S_p(\tau, \nu, \phi)>T_r,\:\:\:\,\forall\left(\tau,\nu,\phi\right)\in\Gamma\times\Omega\right\}\,,
	\end{align}
	where $T_r$ is the detection threshold, to be properly optimized, $\Gamma$ is the Doppler-delay grid described in \ref{subsec:OTFS-Input-Output} and $\Omega$ is defined as a discretized set of angles.\footnote{For example, with an angular sector covering of $60$ degrees divided in $4$ equally spaced parts, the set of angles result to be (supposing the center of the beam to be at $0$ degree) $\Omega=\left\{-30,-15,0,15,30\right\}$.} 
	%such that the illuminated angular sector is equally split.
	If the detection is correct, $\left|\Pc\right|=P$% , where $\left|\cdot\right|$ denotes the number of elements (cardinality) of the set
	, and each target is associated to a coarse estimation $(\hat{\phi}_p,\hat{\tau}_p,\hat{\nu}_p)$.
	\item \textbf{\textit{Super-Resolution Estimation of Radar Parameters for Detected Targets:}}
	\begin{enumerate}[label=2.\arabic*)]
		\item \textit{Fine AOA}: For each detected target, compute
		\begin{align}\label{eq:Phi-Fine}
			\hat{\phi}_p=\arg\max_{\phi}S_p(\hat{\tau}_p,\hat{\nu}_p,\phi),\:\:\:p=1,\dots,\left|\Pc\right|\,.
		\end{align} 
		\item \textit{Fine Doppler-delay Estimation}:
		\begin{enumerate}
			\item \textit{Initialization:} Iteration $i=0$, initialize $\hat{h}'_p[0] = 0$.% for all $p = 1,\dots,\left|\Pc\right|$. 
			\item For iteration $i = 1, 2, 3, \ldots$ repeat: 
			\begin{itemize}
				\item \textit{Delay and Doppler update}:  For each $p = 1,\dots,\left|\Pc\right|$, find the estimates $\hat{\tau}_p[i],\hat{\nu}_p[i]$ 
				by solving the two-dimensional maximization 
				\begin{align}\label{eq:Argmax-Sp-Ip}
					%(\hat{\tau}_p[i],\hat{\nu}_p[i]) & = \arg\max_{\left(\tau_p,\nu_p\right)} \Big \{ S\left(\tau_p,\nu_p, \hat{\phi}_p\right) - \nonumber\\
					%&I_p\left(\{\hat{h}'_q[i-1]\}_{q\neq p},\tau_p,\nu_p,\hat{\phi}_p,  \{ \hat{\tau}_q[i-1],\hat{\nu}_q[i-1]\}_{q\neq p}\right) \Big \}\,.
					(\hat{\tau}_p[i],\hat{\nu}_p[i]) & = \arg\max_{\left(\tau,\nu\right)} \Big \{ S_p - I_p \Big \}\,,
				\end{align}
				with $S_p$ and  $I_p$ computed for $(\hat{h}'_p[i],\tau,\nu, \hat{\phi}_p[i])$.
				\item \textit{Complex channel coefficients update}: Solve the linear system (\ref{eq:459}) using channel matrices $\Gm_p$ with parameters $(\hat{h}'_p[i],\hat{\tau}_p[i],\hat{\nu}_p[i], \hat{\phi}_p)$, and let the solution
				be denoted by $\hat{h}'_p[i]$.
			\end{itemize}
		\end{enumerate}
	\end{enumerate}
	\item \textit{Re-Fine AOA}: Compute \eqref{eq:Phi-Fine} using the refined estimation $\left(\hat{\tau}_p,\hat{\nu}_p\right)$ obtained in \eqref{eq:Argmax-Sp-Ip}.
\end{enumerate}

\begin{remark}
	Since $S_p$ is a convex function in $\phi$ for a fixed pair $(\tau_p,\nu_p)$, the result of \eqref{eq:Phi-Fine} can be exactly computed using common convex solvers. Therefore, the angle can be estimated with super resolution far beyond the discrete grid $\Omega$.
\end{remark}
\begin{remark}
	The detection threshold $T_r$ has been heuristically defined by taking the average of the first $4$ local maxima of the angle-Doppler-delay grid carrying the maximum value of $S_p(\tau, \nu, \phi)$. Simulation results show that this choice is reasonable.
\end{remark}

\subsection{Computational Complexity}
Equation \eqref{eq:3D-Search} describes a method requiring the search over a three dimensional structure composed of $|\Omega|$ slices of $N\times M$ Doppler-delay grids. This search results to be computationally feasible by keeping the dimensions of the sets limited, i.e., considering the Doppler-delay grid %$\Gamma$ (and not making use of a $\Gamma$ finer than the Doppler-delay grid),
and coarse $\Omega$ (angle step). Even if this assumption is rather restrictive, simulation results show that multiple targets can be detected if they are separated at least in one domain out of three (delay, Doppler, angle).

Moreover, note that for fixed $\Gamma$ and $\Omega$, the $|\Omega|\cdot|\Gamma|$ possible different matrices $\Gm_p(\tau_p, \nu_p,\phi_p)$, for $\left(\tau_p,\nu_p\right)\in\Gamma$ and $\phi_p\in\Omega$, do not change, and can be computed once and stored at the receiver. Furthermore, by supposing to send always the same block of symbols (at least during detection when the presence of a target is not given for granted), also the product $\Gm_p\xv$ can be computed once and stored, remarkably simplifying the computationally complexity of the three dimensional search. 

As a final remark, note that during the iterations of the fine Doppler-delay estimation in step 2.2), it is not necessary to search over the entire Doppler-delay grid, but the maximization of the parameters can be restricted, step-by-step, around the previous estimated values. In this way, there is an overall reduction of the algorithmic complexity and feasibility, together with an increase of the accuracy (the estimation step can be reduced if the search is confined in a smaller interval).
	
\subsection{Cram\'er-Rao Lower Bound (CRLB)}\label{sec:Cramer-Rao-Bound}
\newcommand{\Pbar}{ \bar{\Psi}}
We consider the \ac{CRLB} as a theoretical benchmark. In order to estimate a complex channel coefficient, we let $A_p = |h'_p|$ and $\psi_p = \angle(h'_p)$ denote the amplitude and the phase of $h'_p$, respectively. Thus, $5P$ real variables have to be estimated, i.e., $	\thetav =\{A_p, \psi_p, \tau_p, \nu_p,\phi_p\}$. We form the $5P\times 5P$ Fisher information matrix whose $(i,j)$ element is
\begin{align}
[\Id(\thetav) ]_{i,j} =
% \frac{2}{N_0} \Re\left\{\left[\frac{\partial \muv}{\partial \theta_i}\right]^\H \left[\frac{\partial \muv}{\partial \theta_j}\right]\right\} =
 \frac{2}{N_0} \Re\left\{ \sum_{n, m, t} \left[\frac{\partial s_p^{[n ,m, t]}}{\partial \theta_i}\right]^* \left[\frac{\partial s_q^{[n, m,t]}}{\partial \theta_j}\right]\right\}\,, 
\end{align}
where $p = [i]_P$, $q= [j]_P$, and
\begin{align}
	s_p^{[n,m,t]} &= A_p e^{j \psi_p} b_t(\phi_p)a_t^*(\phi_p)f_t \sum_{k=0}^{L-1} \sum_{l=0}^{M-1} \Psi^p_{n, k}[ m, l] x_{k, l}\,,
\end{align}
where $(n,m,t)$ denote time, subcarrier, and antenna, respectively. The desired \ac{CRLB} follows by filling the Fisher information matrix with the corresponding derivatives and obtaining the diagonal elements of the inverse Fisher information matrix.

\section{Simulation Results}\label{sec:Sim-Results}

\newcommand{\scaleVal}{0.13}
\begin{figure}[t!]
	\,
	\begin{subfigure}[b]{0.5\columnwidth}
		\centering
		\includegraphics[scale=\scaleVal]{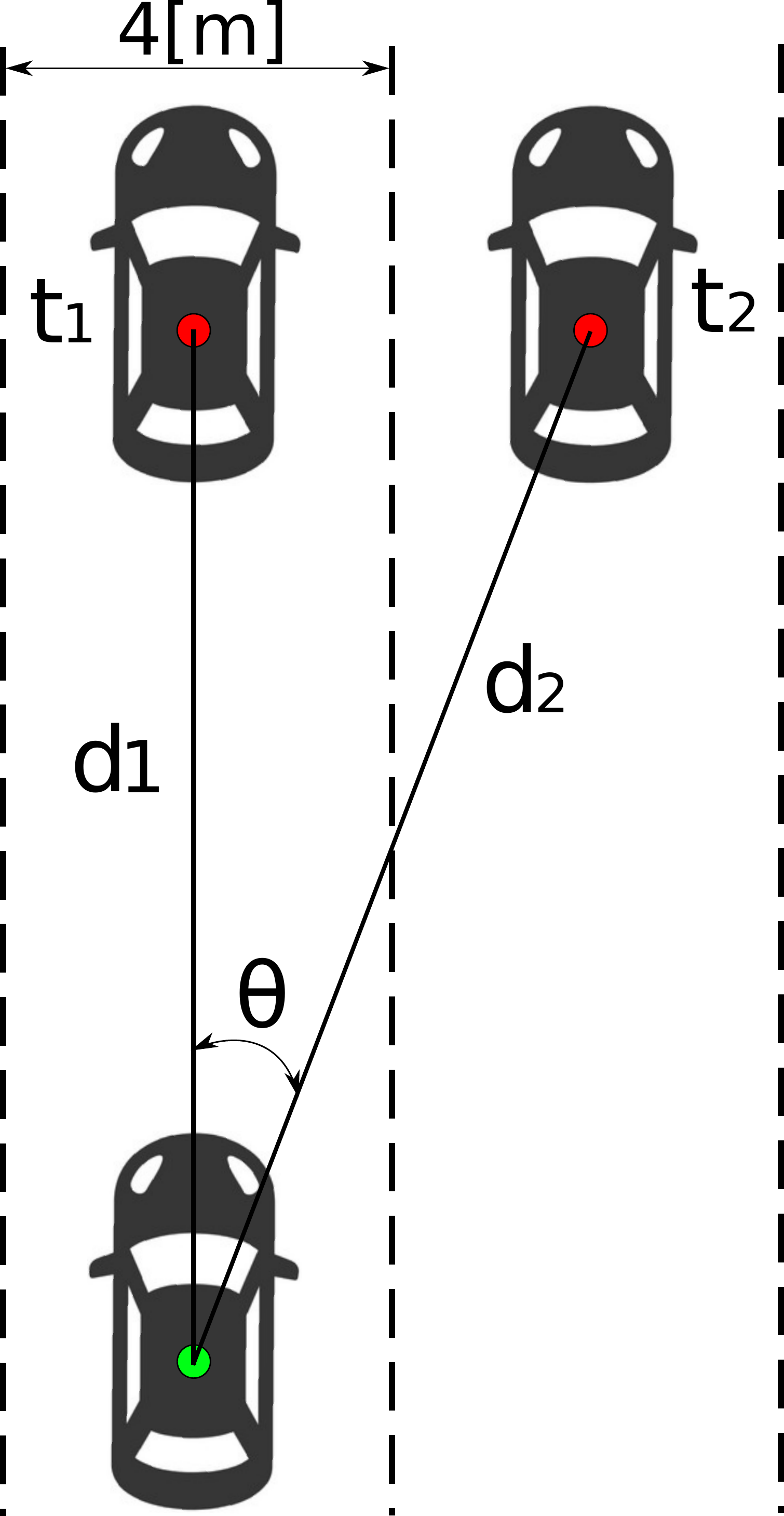}
		\caption{}
	\end{subfigure}%
	\begin{subfigure}[b]{0.5\columnwidth}
		\centering
		\includegraphics[scale=\scaleVal]{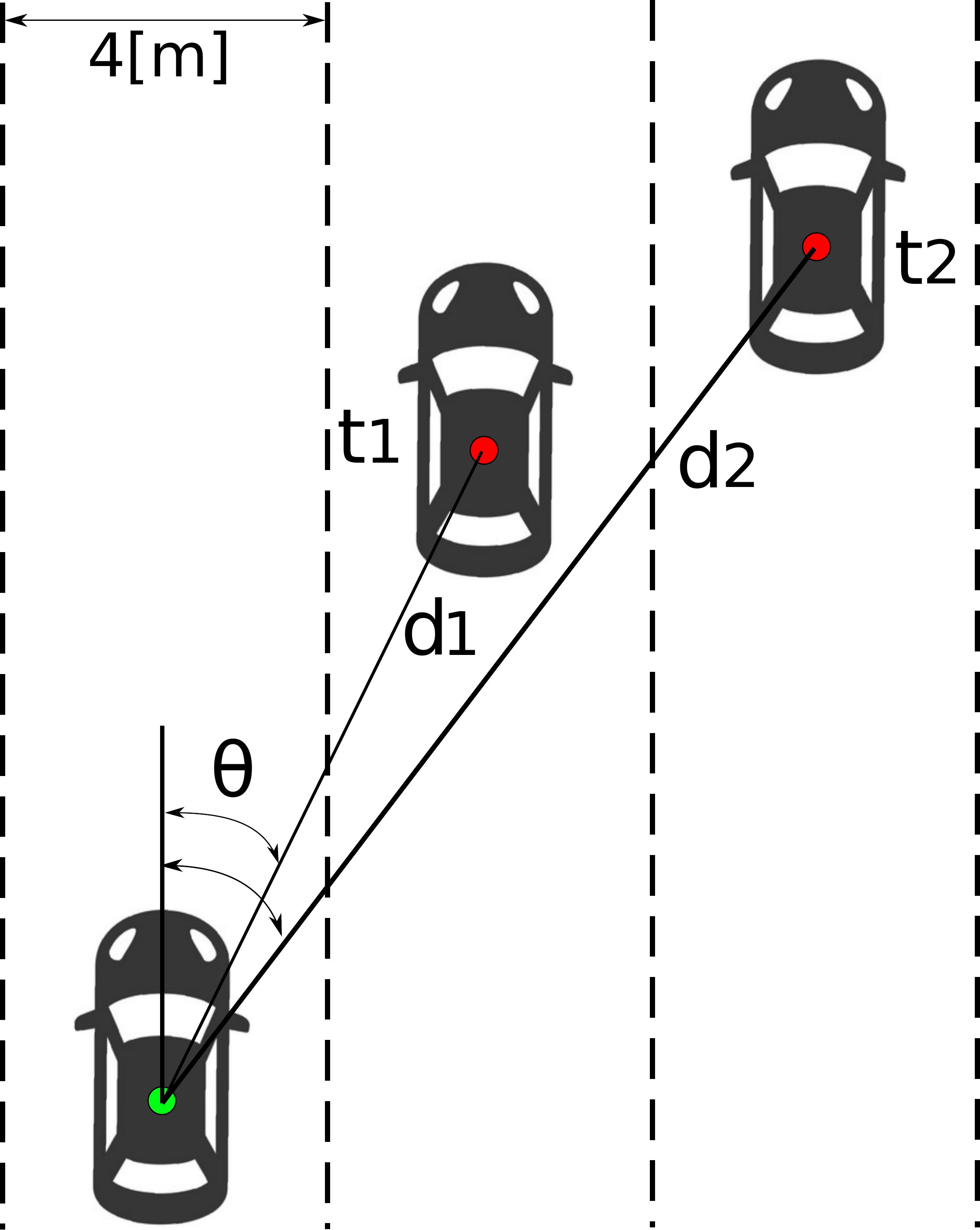}
		\caption{}
	\end{subfigure}%
	%
%	\begin{subfigure}[b]{0.3\columnwidth}
%		\centering
%		\includegraphics[scale=\scaleVal]{scheme3}
%		\caption{}
%	\end{subfigure}
	\caption{Considered automotive scenarios. Green circle: radar transmitter/receiver. Red circles: possible targets. Scenario (a) ($S_{(a)}$): two targets with similar ranges but different angles. Scenario (b) ($S_{(b)}$): two targets with similar angle but different ranges. The line width of a typical highway has been set to $4$ m. Other parameters are listed in Table \ref{tab:System-Parameters}.}
	\label{fig:Scenarios}
\end{figure}

\renewcommand{\arraystretch}{1.2}
\begin{table}
	\caption{System parameters}
	\centering
	\begin{tabular}{|c|c|}
		\hline
		$N=50$ & $M=64$ \\ \hline
		$f_c=60$ [GHz] & $B=150$ [MHz] \\ \hline
		$v_{\mathrm{res}}\simeq421$ [km/h] & $r_{\mathrm{res}}\simeq1$ [m]\\ \hline
		$v_{\mathrm{max}}=N\cdot v_{\mathrm{res}}$ & $r_{\mathrm{max}}=M\cdot r_{\mathrm{res}}$ \\ \hline\hline
		%		\multicolumn{2}{|c|}{\textbf{Scheme (a)}} \\ \hline
		%		$d_1=20$ [m] & $d_2=20.1$ [m] \\ \hline
		%		$v_1=80.2$ [km/h] & $v_2=82.5$ [km/h] \\ \hline
		%		$\theta_1=0.3^{\circ}$ & $\theta_2=6.31^{\circ}$ \\ \hline\hline
		%		\multicolumn{2}{|c|}{\textbf{Scheme (b)}} \\ \hline
		%		$d_1=20.1$ [m] & $d_2=40.2$ [m] \\ \hline
		%		$v_1=80.2$ [km/h] & $v_2=85.5$ [km/h] \\ \hline
		%		$\theta_1=7.2^{\circ}$ & $\theta_2=7.2^{\circ}$ \\ \hline%\hline
	\end{tabular}
	%	\\
	%	\begin{tabular}{|c|c|c|}
	%		\hline
	%		\multicolumn{3}{|c|}{\textbf{Scheme (c)}} \\ \hline
	%		$d_1=20$ [m] & $d_2=20.1$ [m] & $d_3=50.4$ [m] \\ \hline
	%		$v_1=80.2$ [km/h] & $v_2=60.3$ [km/h] & $v_2=122.5$ [km/h] \\ \hline
	%		$\theta_1=0.23^{\circ}$ & $\theta_2=8.87^{\circ}$ & $\theta_2=0.23^{\circ}$ \\ \hline
	%	\end{tabular}
	\\
	\begin{tabular}{|c|c|c|c|c|c|c|}
		\hline
		\textbf{Scheme} & $d_1$ [m] & $d_2$ [m] & $v_1$ [km/h] & $v_2$ [km/h] & $\theta_1^{\circ}$ & $\theta_2^{\circ}$ \\ \hline
		$S_{(a)}$ & 20 & 20.1 & 80.2 & 82.5 & 0.3 & 6.31 \\ \hline
		$S_{(b)}$ & 20.1 & 40.2 & 78.4 & 85.5 & 7.2 & 8.0 \\ \hline
	\end{tabular}
	\label{tab:System-Parameters}
\end{table}

\begin{figure}
	\centering
	\includegraphics[scale=0.65]{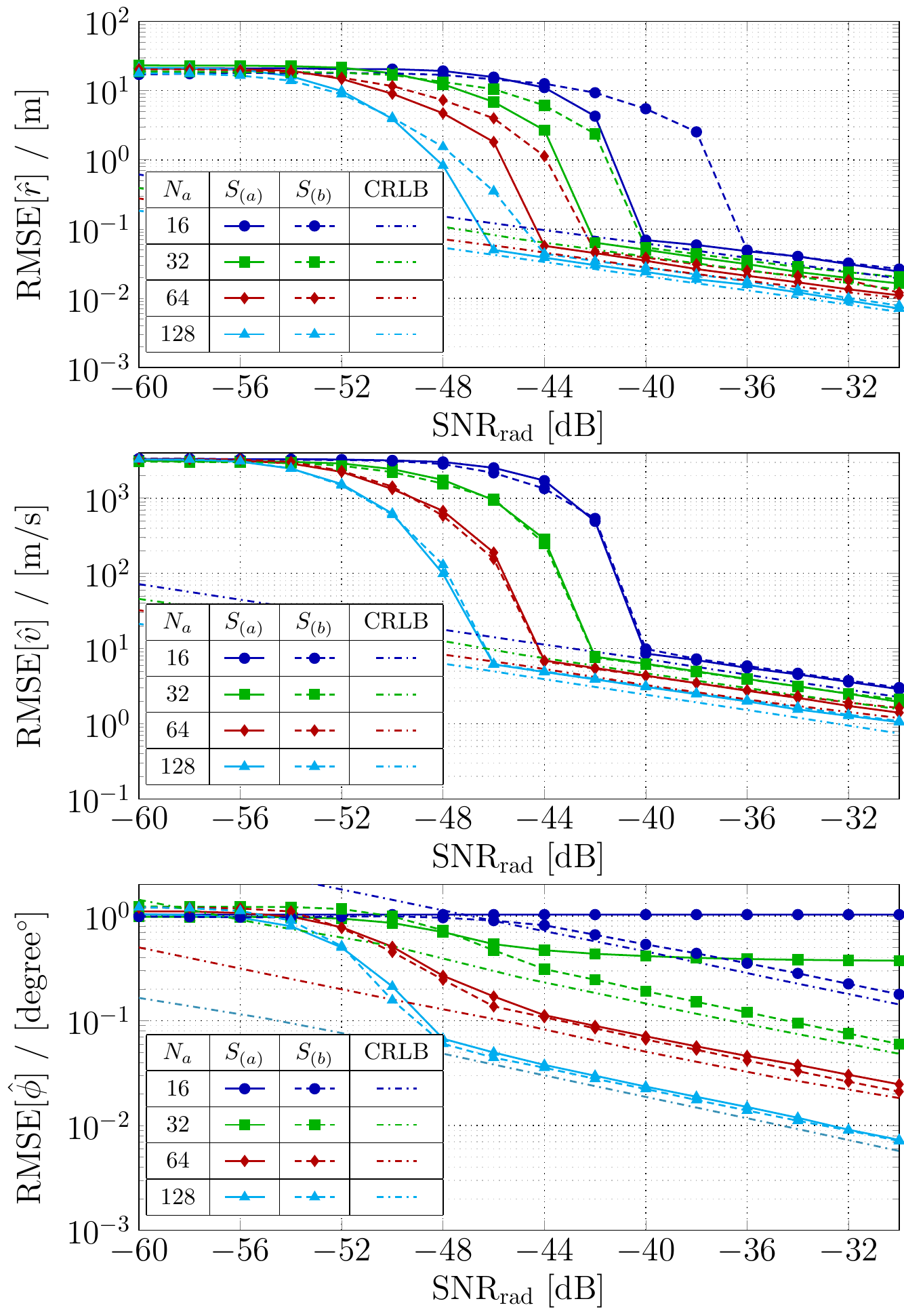}
	\caption{RMSE performance of Scheme (a) ($S_{(a)}$) and Scheme (b) ($S_{(b)}$) of Fig. \ref{fig:Scenarios} with a different number of antennas $\Na$.}
	\label{fig:Ang30-BF}
\end{figure}

\begin{figure}
	\centering
	\includegraphics[scale=0.65]{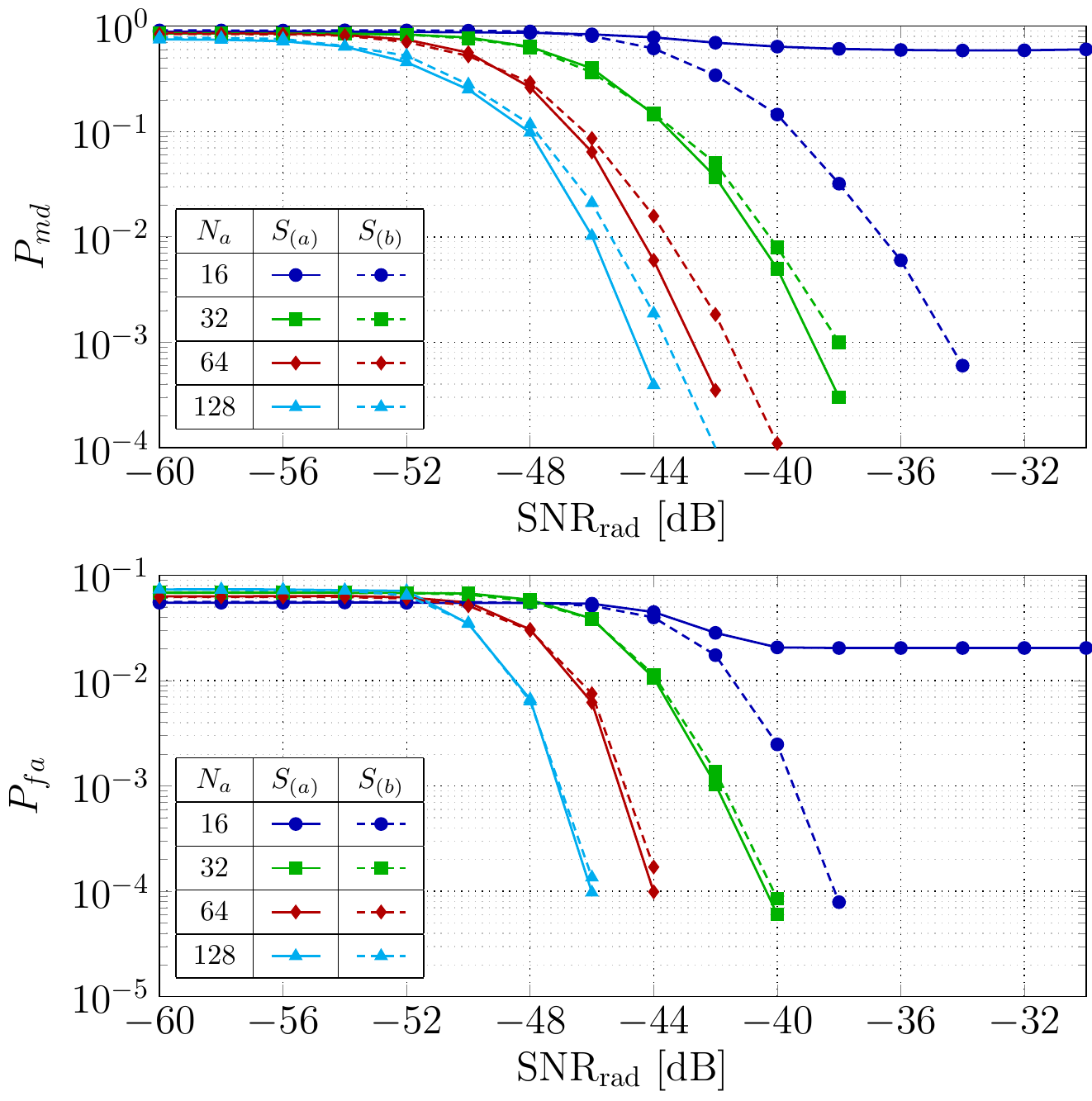}
	\caption{Miss detection $P_{md}$ and false alarm $P_{fa}$ probabilities of Scheme (a) ($S_{(a)}$) and Scheme (b) ($S_{(b)}$) of Fig. \ref{fig:Scenarios} with a different number of antennas $\Na$.}
	\label{fig:Pmd-Pfa}
\end{figure}

%In order to cope with all possible different configurations which might appear in a general automotive scenario, 
In order to cope with some illustrative automotive scenarios, we consider scenarios (a) and (b) of Fig. \ref{fig:Scenarios}, while hybrid scenarios, as a combination of (a) and (b), can be a straightforward extension. Although (a) and (b) are \ac{LoS} scenarios to ensure reliable communication at \ac{mmWave} frequency bands (as in our simulations), the proposed algorithm can be applied to any other scenario and setup including the cases when multiple targets have exactly the same angle (e.g., at lower frequencies). 
Table \ref{tab:System-Parameters} provides the system parameters, together with the parameters of the different analyzed scenarios.\footnote{Note that ranges, velocities, and angles have been chosen randomly and the obtained results are completely independent of these choices.} 
We define the radar \ac{SNR}, from the backscattered signal, as \cite[Chapter 2]{richards2014fundamentals}
\begin{equation}
	\mathrm{SNR_{rad}}=\frac{\lambda^2\sigma_{\mathrm{rcs}}G^2}{\left(4\pi\right)^3r^4}\frac{\Pav}{\sigma_w^2}\,,
\end{equation}
where $\lambda=c/f_c$ is the wavelength, $c$ is the speed of light, $\sigma_{\mathrm{rcs}}$ is the radar cross-section of the target in $\mathrm{m}^2$ ($\sigma_{\mathrm{rcs}}=1$ in our setup), $G$ is the antenna gain ($G=1$ in our setup), $r$ is the distance between transmitter and receiver, and $\sigma_w^2$ is the variance of the \ac{AWGN} noise. In the case of multiple targets with different ranges, we set $\mathrm{SNR_{rad}}$ as the \ac{SNR} of the nearest target.

While two distinct targets in the angle domain can be identified if the angular resolution meets some conditions (depending on the number of antennas, the angular distance between the two targets, and the array properties), a different analysis has to be done for Doppler and delay. In brief, given the parameters defined in Sec. \ref{sec:phy-model}, velocity and range resolutions are
\begin{equation}
v_{\mathrm{res}}=\frac{Bc}{2NMf_c}\ \mathrm{[m/s]}\,,\:\:\:r_{\mathrm{res}}=\frac{c}{2B}\ \mathrm{[m]}\,.
\end{equation}
In order to get a reasonable range resolution, e.g., $<1$ [m], a large bandwidth has to be considered.\footnote{Note that a tradeoff appears. Larger bandwidths mean more precise resolution, but lower maximum range (with the same $N\times M$ grid). We remark that our algorithm is completely independent of these choices.} Since the velocity resolution is directly proportional to $B$, for a fixed $f_c$, the only way to obtain a fine resolution is to increase the block size $NM$, leading to a remarkable increase in computational complexity, which could be not affordable. For this reason, given the fact that multiple targets can be detected if separable in at least one domain out of three (angle, Doppler, delay), we fix the system parameters by focusing on a reasonable range resolution (and maximum range) under a feasible computational complexity. This in turn leads to an unavoidable very large velocity resolution. Clearly, under the system parameters of Table \ref{tab:System-Parameters}, multiple targets are not separable solely in the Doppler domain.

\begin{remark}
	The performance of the proposed \ac{ML}-based algorithm strictly depends on the dimension of the block of data sent, i.e., the product $N\cdot M$. Thus, the system parameters of Tab. \ref{tab:System-Parameters} can be easily tuned to achieve the desired levels of radar resolutions (modifying the bandwidth), acquisition time (based on the length of the \ac{OTFS} frame in time), maximum range, etc. Clearly, the \ac{CRLB} changes accordly. Moreover, note that this is also possible thanks to \ac{OTFS} modulation, which is not sensitive to Doppler and delay effects.   
\end{remark}

%\subsection{Numerical Results}
Fig. \ref{fig:Ang30-BF} shows the performance in terms of \ac{RMSE} when an angular sector of $30$ degrees (from $-15^\circ$ to $15^\circ$) is covered (through an appropriate definition of the \ac{BF} vector $\fv_{\mathrm{BF}}$) for both scenarios (a) ($S_{(a)}$) and (b) ($S_{(b)}$) of Fig. \ref{fig:Scenarios}, with parameters listed in Tab. \ref{tab:System-Parameters}. Note that the \ac{RMSE} is computed given the detection of the target within the angular sector. However, even if a target is correctly detected, if the \ac{SNR} is not high enough, the parameter estimation performance might be not satisfactory. The probabilities of miss detection and false alarm, denoted by $P_{md}$ and $P_{fa}$, are provided in Fig. \ref{fig:Pmd-Pfa}. The miss detection occurs when the presence of a target is not detected, i.e., the backscattered power is below the threshold, while the false alarm occurs when the received power in an angular sector is higher than the threshold, but no target is present.

In Fig. \ref{fig:Ang30-BF}, we notice that after the ``waterfall'' transition,  typical of \ac{ML} estimators, all estimators achieve the respective \ac{CRLB} up to a small-to-negligible loss due to the discretization error of the search over the three domains. In scenario (a), the algorithm is not able to distinguish two targets closed each other in the angle domain with $\Na=16$ antennas as observed in Fig. \ref{fig:Pmd-Pfa}. This results in very poor angle estimation as seen in Fig. \ref{fig:Ang30-BF}. This behavior does not occur in scenario (b), even with $\Na=16$, as two targets are well separated in the range domain.
With $\Na=32$ antennas, the algorithm is able to distinguish two targets in scenario (a), as seen in Fig. \ref{fig:Pmd-Pfa}, but cannot still achieve the desired angle estimation performance.  Overall, the increase of the number of antennas leads to better asymptotic \ac{RMSE} performance (lower \ac{CRLB}) and detection probabilities, together with the capability of identifying nearest targets in the angle domain and reach the \ac{CRLB} in any scenario.

Moreover, note that the heuristic definition of the detection threshold in Sec. \ref{sec:Joint-Detection-Param-Estimation} provides reasonable performance. In fact, beyond threshold \ac{SNR} values at which the \ac{RMSE} curves reach the corresponding \ac{CRLB}, both probabilities of miss detection and false alarm achieve remarkable performance. Note that it is meaningful to operate at \ac{SNR} values much higher than the ``waterfall'' behavior so that desired parameter estimation can be achieved.  

\section{Conclusion}\label{sec:Conclusion}
%In this paper, we analyzed the radar target detection performance of a joint radar and communication system based on OTFS modulation. Simulation results show that multiple targets are well identified if separable in at least one domain out of three (delay, Doppler, or angle), while achieving a small-to-negligible loss in terms of estimation error of typical radar parameters w.r.t. the theoretical benchmark, i.e., the \ac{CRLB}. Moreover, the proposed algorithm, based on a \ac{ML} approach, is fully independent on the system setup, not precluding its use in a particular scenario. 
We studied the joint target detection and parameter estimation in a simple setup where a transmitter equipped with mono-static MIMO radar wishes to detect multiple targets and estimate the corresponding parameters using OTFS modulation and wide sector beams. We proposed an efficient \ac{ML}-based approach that achieves good tradeoff between detection and estimation performance and complexity. Simulation results show that multiple targets are well identified if separable in at least one domain out of three (delay, Doppler, or angle), while achieving a small-to-negligible loss in terms of estimation error of typical radar parameters compared to \ac{CRLB}. Moreover, the proposed algorithm is fully independent on the system setup and not restricted to any particular scenario. 
\bibliography{IEEEabrv,book}

\begin{thebibliography}{10}
\providecommand{\url}[1]{#1}
\csname url@samestyle\endcsname
\providecommand{\newblock}{\relax}
\providecommand{\bibinfo}[2]{#2}
\providecommand{\BIBentrySTDinterwordspacing}{\spaceskip=0pt\relax}
\providecommand{\BIBentryALTinterwordstretchfactor}{4}
\providecommand{\BIBentryALTinterwordspacing}{\spaceskip=\fontdimen2\font plus
\BIBentryALTinterwordstretchfactor\fontdimen3\font minus
  \fontdimen4\font\relax}
\providecommand{\BIBforeignlanguage}[2]{{%
\expandafter\ifx\csname l@#1\endcsname\relax
\typeout{** WARNING: IEEEtran.bst: No hyphenation pattern has been}%
\typeout{** loaded for the language `#1'. Using the pattern for}%
\typeout{** the default language instead.}%
\else
\language=\csname l@#1\endcsname
\fi
#2}}
\providecommand{\BIBdecl}{\relax}
\BIBdecl

\bibitem{li2008mimo}
J.~Li and P.~Stoica, \emph{{MIMO radar signal processing}}.\hskip 1em plus
  0.5em minus 0.4em\relax John Wiley \& Sons, 2008.

\bibitem{patole2017automotive}
S.~M. {Patole}, M.~{Torlak}, D.~{Wang}, and M.~{Ali}, ``Automotive radars: A
  review of signal processing techniques,'' \emph{{IEEE} Signal Process. Mag.},
  vol.~34, no.~2, pp. 22--35, March 2017.

\bibitem{song2018scalable}
X.~Song, S.~Haghighatshoar, and G.~Caire, ``A scalable and statistically robust
  beam alignment technique for millimeter-wave systems,'' \emph{{IEEE} Trans.
  Wireless Commun.}, vol.~17, no.~7, pp. 4792--4805, 2018.

\bibitem{friedlander2012transmit}
B.~Friedlander, ``On transmit beamforming for {MIMO} radar,'' \emph{{IEEE}
  Trans. Aerosp. Electron. Syst.}, vol.~48, no.~4, pp. 3376--3388, 2012.

\bibitem{niesen2019joint}
U.~Niesen and J.~Unnikrishnan, ``Joint beamforming and association design for
  {MIMO} radar,'' \emph{{IEEE} Trans. Signal Process.}, vol.~67, no.~14, pp.
  3663--3675, 2019.

\bibitem{richards2014fundamentals}
M.~A. Richards, \emph{Fundamentals of radar signal processing, Second
  edition}.\hskip 1em plus 0.5em minus 0.4em\relax McGraw-Hill Education, 2014.

\bibitem{bar1995multitarget}
Y.~Bar-Shalom and X.-R. Li, \emph{Multitarget-multisensor tracking: principles
  and techniques}.\hskip 1em plus 0.5em minus 0.4em\relax YBs Storrs, CT, 1995,
  vol.~19.

\bibitem{gaudio2019effectiveness}
L.~Gaudio, M.~Kobayashi, G.~Caire, and G.~Colavolpe, ``On the effectiveness of
  {OTFS} for joint radar and communication,'' \emph{arXiv preprint
  arXiv:1910.01896}, 2019.

\bibitem{hadani2017orthogonal}
R.~Hadani, S.~Rakib, M.~Tsatsanis, A.~Monk, A.~J. Goldsmith, A.~F. Molisch, and
  R.~Calderbank, ``Orthogonal time frequency space modulation,'' in \emph{2017
  IEEE Wireless Commun. and Network. Conf. (WCNC)}.\hskip 1em plus 0.5em minus
  0.4em\relax IEEE, 2017, pp. 1--6.

\bibitem{dokhanchi2019mmwave}
S.~H. Dokhanchi, B.~S. Mysore, K.~V. Mishra, and B.~Ottersten, ``{A mmWave
  automotive joint radar-communications system},'' \emph{{IEEE} Trans. Aerosp.
  Electron. Syst.}, vol.~55, no.~3, pp. 1241--1260, June 2019.

\bibitem{zheng2019radar}
L.~Zheng, M.~Lops, Y.~C. Eldar, and X.~Wang, ``Radar and communication
  co-existence: an overview: a review of recent methods,'' \emph{{IEEE} Signal
  Process. Mag.}, vol.~36, no.~5, pp. 85--99, Sep. 2019.

\bibitem{hassanien2019dual}
A.~{Hassanien}, M.~G. {Amin}, E.~{Aboutanios}, and B.~{Himed}, ``Dual-function
  radar communication systems: A solution to the spectrum congestion problem,''
  \emph{{IEEE} Signal Process. Mag.}, vol.~36, no.~5, pp. 115--126, Sep. 2019.

\bibitem{raviteja2019orthogonal}
P.~Raviteja, K.~T. Phan, Y.~Hong, and E.~Viterbo, ``Orthogonal time frequency
  space ({OTFS}) modulation based radar system,'' in \emph{2019 IEEE Radar
  Conf. (RadarConf)}.\hskip 1em plus 0.5em minus 0.4em\relax IEEE, 2019, pp.
  1--6.

\bibitem{sabharwal2014band}
A.~{Sabharwal}, P.~{Schniter}, D.~{Guo}, D.~W. {Bliss}, S.~{Rangarajan}, and
  R.~{Wichman}, ``In-band full-duplex wireless: Challenges and opportunities,''
  \emph{{IEEE} J. Sel. Areas Commun.}, vol.~32, no.~9, pp. 1637--1652, Sep.
  2014.

\bibitem{song2019fully}
X.~Song, T.~K{\"u}hne, and G.~Caire, ``Fully-/partially-connected hybrid
  beamforming architectures for {mmWave} {MU-MIMO},'' \emph{{IEEE} Trans.
  Wireless Commun.}, vol.~19, no.~3, pp. 1754--1769, 2020.

\bibitem{kumari2018ieee}
P.~{Kumari}, J.~{Choi}, N.~{González-Prelcic}, and R.~W. {Heath}, ``{IEEE}
  802.11ad-based radar: An approach to joint vehicular communication-radar
  system,'' \emph{{IEEE} Trans. Veh. Technol.}, vol.~67, no.~4, pp. 3012--3027,
  April 2018.

\bibitem{nguyen2017delay}
D.~H.~N. {Nguyen} and R.~W. {Heath}, ``{Delay and Doppler processing for
  multi-target detection with IEEE 802.11 OFDM signaling},'' in \emph{Proc.
  {IEEE} Int. Conf. Acoustics, Speech, and Signal Processing {(ICASSP)}}, March
  2017, pp. 3414--3418.

\bibitem{vitetta2013wireless}
G.~A. Vitetta, D.~P. Taylor, G.~Colavolpe, F.~Pancaldi, and P.~A. Martin,
  \emph{Wireless communications: algorithmic techniques}.\hskip 1em plus 0.5em
  minus 0.4em\relax John Wiley \& Sons, 2013.

\bibitem{fortunati2020mimoRadar}
S.~{Fortunati}, L.~{Sanguinetti}, F.~{Gini}, M.~S. {Greco}, and B.~{Himed},
  ``Massive {MIMO} radar for target detection,'' \emph{{IEEE} Trans. Signal
  Process.}, vol.~68, pp. 859--871, 2020.

\bibitem{matz2013time}
G.~{Matz}, H.~{Bolcskei}, and F.~{Hlawatsch}, ``Time-frequency foundations of
  communications: Concepts and tools,'' \emph{{IEEE} Signal Process. Mag.},
  vol.~30, no.~6, pp. 87--96, Nov 2013.

\end{thebibliography}

\end{document}